\documentclass[%
reprint,
superscriptaddress,
 amsmath,amssymb,
 aps,
floatfix,
]{revtex4-2}

\usepackage{graphicx}
\usepackage{dcolumn}
\usepackage{bm}
\usepackage{appendix}

\usepackage{color}
\usepackage{physics}

\begin{document}
\preprint{APS/123-QED}

\title{Transverse spin angular momentum of space-time surface plasmon polariton wave packet}
\author{Naoki Ichiji}
\affiliation{Graduate School of Pure and Applied Sciences, University of Tsukuba, 1-1-1 Tennodai, Tsukuba-shi, Ibaraki 305-8571, Japan}
\author{Daigo Oue}
\affiliation{Instituto Superior T\'{e}cnico, University of Lisbon, 1049-001 Lisbon, Portugal}
\affiliation{Department of Physics, Kyoto University, Kyoto 606-8502, Japan}
\affiliation{The Blackett Laboratory, Imperial College London, London SW7 2AZ, United Kingdom}
\author{Murat Yessenov}
\affiliation{CREOL, The College of Optics \& Photonics, University of Central Florida, Orlando, FL 32816, USA}
\author{Kenneth L. Schepler}
\affiliation{CREOL, The College of Optics \& Photonics, University of Central Florida, Orlando, FL 32816, USA}
\author{Ayman F. Abouraddy}
\affiliation{CREOL, The College of Optics \& Photonics, University of Central Florida, Orlando, FL 32816, USA}
\author{Atsushi Kubo}
\thanks{kubo.atsushi.ka@u.tsukuba.ac.jp}
\affiliation{Faculty of Pure and Applied Sciences, University of Tsukuba,1-1-1 Tennodai, Tsukuba-shi, Ibaraki 305-8571, Japan}

\begin{abstract}
In addition to longitudinal spin angular momentum (SAM) along the axis of propagation of light, spatially structured electromagnetic fields such as evanescent waves and focused beams have recently been found to possess transverse SAM in the direction perpendicular to the axis of propagation. 
In particular, the SAM of SPPs with spatial structure has been extensively studied in the last decade after it became clear that evanescent fields with spatially structured energy flow generate three-dimensional spin texture. 
Here we present numerical calculations of the space-time surface plasmon polariton (ST-SPP) wave packet, a plasmonic bullet that propagates at an arbitrary group velocity while maintaining its spatial distribution. ST-SPP wave packets with complex spatial structure and energy flow density distribution determined by the group velocity are found to propagate with accompanying three-dimensional spin texture and finite topological charge density.
Furthermore, the spatial distribution of the spin texture and topological charge density determined by the spatial structure of the SPP is controllable, and the deformation associated with propagation is negligible. ST-SPP wave packets, which can stably transport customizable three-dimensional spin textures and topological charge densities, can be excellent subjects of observation in studies of spinphotonics and optical topological materials.
\end{abstract}

\maketitle
\onecolumngrid
\section{Introduction}
Spin angular momentum (SAM) associated with the rotation of electromagnetic polarization~\cite{poynting1909wave,beth1935direct,beth1936mechanical} plays an important role in both classical and quantum optics~\cite{sugiura1990optical,higurashi1994optically,higurashi1997optically,friese1998optical,ahn2018optically,monteiro2018optical,reimann2018ghz,jin20216}. While the SAMs of light have been studied for nearly a century, most discussions until the 2000s focused on the longitudinal SAM ($l$-SAM), which is the component parallel to the propagation axis. However, Aiello and Bliokh et al.~reported that optical SAM also has a component perpendicular to the propagation axis, transverse SAM ($t$-SAM)~\cite{Aiello09PRL, Aiello10PRA, Bliokh15PRR}.
$t$-SAM has recently received much attention as it is responsible for a long-standing problem of polarization-dependent beam shift (the Imbert-Fedorov shift)~\cite{bekshaev2009oblique,aiello2009transverse,aiello2010transverse}, has been given new rotational degrees of freedom in optical manipulation~\cite{antognozzi2016direct,canaguier2015plasmonic,canaguier2014transverse} and provides strong light-matter coupling, which cannot be achieved by ordinary polarizations~\cite{junge2013strong}. $t$-SAM is proportional to the rotation of the energy flow density of the electromagnetic field and arises when the Poynting vector of the electromagnetic field has a spatial gradient in the plane normal to the propagation direction of the optical field, such as in tightly focused beams~\cite{Aiello15NP, Zhang22OE}, guided light~\cite{Abujetas20ACSP}, interference beats between two plane weaves~\cite{Nori15PRX}, and evanescent fields~\cite{Nori14NCom}.

Surface plasmon polaritons (SPPs), which are surface electromagnetic waves confined at the metal-insulator interface, are accompanied by an evanescent field that decays in the perpendicular direction to the interface and thus exhibits $t$-SAM in the in-plane direction ($t$-SAM$_{\parallel}$)~\cite{Bliokh12PRA, kim2012time, kim2015spin, Bliokh17PRL, Yanan18ACSN}. The direction of the $t$-SAM$_{\parallel}$ of the SPP is fixed according to the energy propagation direction, resulting in a broken time-reversal symmetry. 
Hence, as a counterpart of a quantum spin Hall effect of light~\cite{Nori15Sci, Mechelen16OPT}, the $t$-SAM$_{\parallel}$ of SPPs has garnered widespread interest in the field of spinphotonics~\cite{Zayats14NC,Zayats13Sci,Yanan19ACSP, Shi21NP, Cardano15NP}.

Furthermore, when the SPP has a two-dimensional spatial structure at the confined interface, its $t$-SAM is not limited to the in-plane component but also includes an out-of-plane component ($t$-SAM$_{\perp}$) ~\cite{Du19NatPhys, Yanan20Nature, Lei21PRL, Zhang21PRR, Ghosh21APR, Yanan22APR, Shi21LPR, Shi21PNAS, Taneja21APL}. The existence of both $t$-SAM$_{\perp}$ and $t$-SAM$_{\parallel}$ components of SPP indicates that the SPP can have a three-dimensional arrangement of spin textures and thus a finite surface area on the Poincar{\'e} polarization vector sphere~\cite{Yanan22NatRP, Ghosh23ACSP, Shen22ACSP, Shen21OL}.
Efforts to generate topologically stable three-dimensional spin vector fields by preparing a vortex-like energy flow have been actively pursued in the context of topological quasiparticles~\cite{Du19NatPhys, Yanan20Nature}. 
The introduction of integer topological charges into SPPs using the designed light-SPP coupler or vortex beam has enabled the demonstration of various plasmonic topological quasiparticles, including plasmonic spin skyrmion~\cite{Du19NatPhys, Zhang21PRR, Deng22NC, Yanan22APR}, topological melon structure~\cite{Yanan20Nature, Xiong21NL}, and topological plasmonic field~\cite{Tsesses18Sci, Davis20Sci} and spin~\cite{Lei21PRL, Ghosh21APR} lattice structures.
However, the spin textures of SPPs discussed in previous studies are typically excited at specific coordinates and do not involve spatial movement, such as propagation. While numerous studies have examined and experimentally realized the motion control of magnetic skyrmions~\cite{Nagaosa13NatN, Jiang15Sci, Woo16NatMat, Jiang17PR, Zhang18NatCom}, only a few have considered the spatial movement of the spin texture of SPPs~\cite{Bai20OE,Lin21ACSP}.

Shi et al.~recently formulated the $t$-SAM of the structured electromagnetic guided waves and experimentally demonstrated the $t$-SAM$_{\perp}$ of SPP waves with inhomogeneous energy flow density, such as Cosine SPP and Airy SPP waves~\cite{Shi21PNAS}.
Their formulation of the SAM suggests that all SPP waves with both spatial structure and energy flow should be accompanied by $t$-SAM$_{\perp}$, even if these SPP fields are topologically uncharged.
Therefore, temporal variation of the SPP fields with spatial structure, such as SPP wave packets with short temporal widths~\cite{Ichiji22NanoP}, steering SPP beams excited by achromatic incident waves~\cite{Wang20OE}, and plasmonic bullets with confinement in all dimensions~\cite{Karalis19PRL}, are expected to be accompanied by $t$-SAM$_{\perp}$ in their propagation.
However, although several observations of $t$-SAM$_{\perp}$ associated with SPPs have been reported~\cite{Taneja21APL}, the migration of $t$-SAM$_{\perp}$ caused by propagating SPP fields has rarely been investigated.

Here, we present theoretical calculations of the $t$-SAM$_\perp$ for the space-time SPP (ST-SPP) wave packet, a newly proposed type of SPP wave packet classified as a plasmonic bullet~\cite{Schepler20ACSP}. ST-SPP wave packets are structured SPPs with artificially designed propagation characteristics achieved through precise adjustments to the spatial and temporal frequencies of each frequency component of the SPP wave packet. The concept of designing propagation characteristics by adjusting the time-space-frequency correlation of a wave packet was initially proposed and experimentally demonstrated for light pulses and referred to as space-time (ST) waves~\cite{Kondakci17NP,Bhaduri18OE,Yessenov22AOP}. 
ST-SPP wave packets have been theoretically shown to possess attractive properties such as diffraction-free and dispersion-free propagation invariance and arbitrary group velocities. Efforts to experimentally excite and observe them are ongoing~\cite{Ichiji22ACSP,Ichiji22arXiv}.

In this study, we conducted an examination of the three-dimensional spin arrangement of ST-SPP wave packets through calculations of each electromagnetic component and energy flow density. Our calculations revealed that ST-SPP wave packets possess a three-dimensional spin texture and a finite topological charge density. The spin texture and topological charge density of ST-SPPs with propagation invariance were found to be spatially robust and maintain a stable spatial distribution during propagation. Interestingly, the topological charge density of ST-SPP wave packets exhibited spatial distributions with only opposite signs on the left and right, with positive and negative positional relationships reversing depending on the group velocity.
The proposed calculation results open up the possibility of constructing plasmonic topological quasiparticles and manipulating excited structured SPP waves, which may have potential applications in optical information transfer and optical trapping.

\section{ST-SPP wave packet}
\begin{figure}[t!h!]
  \begin{center}
  \includegraphics[width=0.7\textwidth]{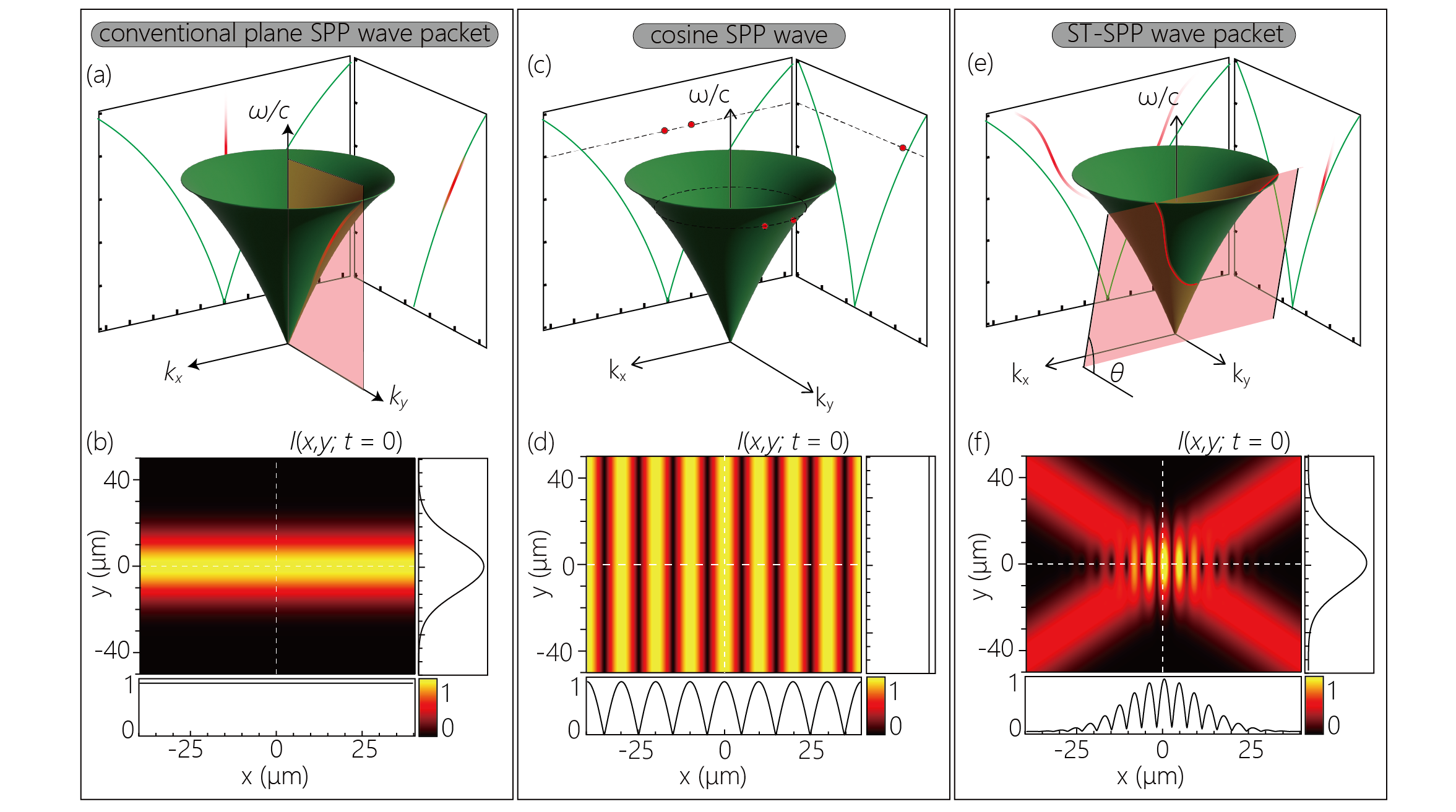}
  \end{center}
  \caption{(a) The spectral representation of conventional plane SPP wave on the surface of the SPP light-cone $k_{x}^2+k_{y}^2=k_{\mathrm{sp}}^2$. (b) Spatial distribution of the field intensity of conventional plane SPP wave in physical space, $I(x,y; t\!=\!0)$. (c,d) Same as (a,b) but for a Cosine SPP wave. (e,f) Same as (a,b) but for an ST-SPP wave. The black lines in the side and bottom panels in (b), (d) and (f) are cross sections through the field distributions in the main panels at $x\!=\!0$ and $y\!=\!0$. Each profile in (b), (d), and (f) is normalized by each maximum value.}
  \label{Fig:Efield}
\end{figure}
To describe the dispersion relation of a wave packet with a two-dimensional spatial distribution, such as an ST-SPP wave packet, a three-dimensional spatiotemporal spectral diagram consisting of one frequency axis and two wavenumber axes is useful.
While the light cone of the light in a vacuum is a cone $k_{\mathrm{l}}\!=\!\omega/c$, the three-dimensional spatiotemporal correlation of SPP, the SPP light cone, has a curved surface that reflects the dispersion of  SPP (Fig.~\ref{Fig:Efield})~\cite{Schepler20ACSP,Lassaline20Nat}, 
\begin{align}
k_{\mathrm{sp}}\!=\! \frac{\omega}{c}\sqrt{\frac{\epsilon_{\mathrm{m}}(\omega)\epsilon_{\mathrm{d}}}{\epsilon_{\mathrm{m}}(\omega)+\epsilon_{\mathrm{d}}}},\label{Eq:SPP-dispersion}
\end{align}
where $\epsilon_{\mathrm{d}(\mathrm{m})}$ is the relative permittivity of the dielectric (metal).

In the case of the conventional pulsed SPP wave with uniform intensity in the transverse direction, the spectral support domain of the SPP wave packet corresponds to the intersection of the SPP-light cone and $k_{x}\!=\!0$ plane (Fig.~\ref{Fig:Efield}(a)). We choose the $y$-direction as the axial propagation direction and define the $z$-direction as the surface normal of the interface between metal and dielectric so that the SPP is localized on the $xy$-plane.
The out-of-plane component of the electric field distribution of the conventional SPP plane wave packet with finite width along the propagation direction (Fig.~\ref{Fig:Efield}(b)) is given by $\int\!\dd{\omega}\;\widetilde{E}(\omega)e^{i(k_{y}y-\omega t)}$, where $\omega$ is the temporal angular frequency and $\widetilde{E}(\omega)$ is the pulse spectrum, the Fourier transform of $E(t)$ at $y\!=\!0$. The free-space pulse used to excite the SPP has the form 
\begin{align}
E(t)\!=\!E_{\mathrm{o}}e^{-i\omega_{\mathrm{o}}t}\exp\left(-\frac{t^{2}}{(\Delta T/(2\ln{2}))^{2}}\right),
\end{align}
where $\Delta T\!=\!100$~fs, $\omega_{\mathrm{o}}/(2\pi)\!=\!375$~THz (800 nm wavelength), and $E_{\mathrm{o}}$ is the amplitude of the electric field. The dielectric is assumed to be vacuum $(\epsilon_{\mathrm{d}}\!=\!1)$ and the metal is gold (Au). We employ the Lorentz-Drude model~\cite{Rakic98AO} for $\epsilon_{\mathrm{m}}(\omega)$.
Experimentally, the conventional plane SPP wave packet corresponds to a wave packet excited by a linear light-SPP coupling structure with sufficiently long length, such as a slit or gratings~\cite{Kubo07NL, Zhang11PRB}. The propagation of a conventional SPP wave on the dispersive sample surface is inevitably accompanied by deformation and chirping due to the frequency dependence of the group velocity~\cite{Ichiji22NanoP}.

As a comparison, the spectral support domain of the Cosine SPP wave~\cite{Lin12PRL} is shown in Fig.~\ref{Fig:Efield}(c). The Cosine SPP waves are a spatial pattern generated by the interference of two SPPs, corresponding to the two-wave interference in free space~\cite{Nori15PRX}. On the SPP light cone surface, the spectral support domain of the Cosine SPP wave is defined by two points that have identical $k_{y}$ components and two $k_{x}$ components equal in magnitude but with opposite signs. The spatial field distribution $E_{z}^{\mathrm{cos}}$ is given by: 
\begin{align}
E_{z}^{\mathrm{cos}}(\mathbf{x}\:|\:\omega)&=\frac{1}{2}\widetilde{E}(\omega)(e^{i(k_{x}x+k_{y}y-\omega t)}+e^{i(-k_{x}x+k_{y}y-\omega t)})\\
&=\widetilde{E}(\omega)\cos (k_{x}x)e^{+i(k_{y}y-\omega t)}
\end{align}
where we have defined $\mathbf{x} = (x,y;t)$. In this study, $E_{z}$ is calculated for the insulator side ($z > 0$).
Note that $k_{x}$ and $k_{y}$ are the real components of the wave vector along the transverse ($x$) and axial ($y$) coordinates, respectively, which satisfy the relation $k_{x}^2+k_{y}^2\!=\!k_{\mathrm{sp}}^2$. The interference between the two plane waves generates a periodic cosine-shaped field distribution in the transverse direction, as its intensity profile is shown in Fig.~\ref{Fig:Efield}(d).

Importantly, the axial wavenumber of the Cosine SPP wave, $k_{y}\!=\!\sqrt{k^2_{\mathrm{sp}}-k_{x}^{2}}$ can be arbitrarily set in the range $0\!<\!k_{y}\!<\!k_{\mathrm{sp}}$ by properly selecting $k_{x}$.
Therefore, by making all frequency components of the SPP wave packet into Cosine SPP waves with designed $k_{y}$, structured SPP wave packets with arbitrary dispersion on the propagation axis can be constructed.
The concept of designing the dispersion relation on the propagation axis by strictly adjusting $k_{x}$ and $k_{y}$ has already been experimentally demonstrated for light pulses in free space and is known as a Space-Time (ST) wave packet~\cite{Kondakci17NP,Bhaduri18OE,Yessenov22AOP}. ST wave packets with a variety of novel propagation characteristics have been reported, including diffraction-free property~\cite{Kondakci17NP}, arbitrary group velocity~\cite{Kondakci19NC}, acceleration and deceleration in unprecedented ranges~\cite{Yessenov20PRL2, Hall22OLaccel}, the introduction of dispersion properties into a light pulse in free space~\cite{Yessenov21ACSP,Hall21PRA}, and non-dispersive propagation in a dispersive media~\cite{He22LPR}.

The spectral support domain of an ST-SPP wave packet, which is the target of this study, is defined as a one-dimensional trajectory at the intersection of the SPP light cone, Eq.~\eqref{Eq:SPP-dispersion}, with a spectral plane 
$\Omega\!=\!(k_{y}-k'_{o})c\;\mathrm{tan}\,\theta$, that is parallel to the $k_{x}$-axis and makes an angle $\theta$ with the $k_{y}$-axis (Fig.~\ref{Fig:Efield}(e)), where $k'_{o}$ is the SPP wave number evaluated at the carrier frequency [i.e., $k' _ o = k _ {\mathrm{sp}}(\omega=\omega_{o})$], and $\Omega\!=\!\omega-\omega_{o}$ is the frequency measured with respect to $\omega_{o}$.

The projection of the spectral support domain onto the $(k_{y},\frac{\omega}{c})$-plane is a straight line, and thus the group velocity $\widetilde{v_{g}}\!=\!\dd{\omega}/dk_{y}\!=\!c\:\mathrm{tan}\,\theta$ is a constant independent of $\omega$. Each component of the spatial field distribution of the ST-SPP wave packet on the $z \!=\! 0^{+}$ plane, $E^{\mathrm{ST}}_{x}$, $E^{\mathrm{ST}}_{y}$ and $E^{\mathrm{ST}}_{z}$ is given by~\cite{Shi21PNAS}(See also appendix for the details):
\begin{align}
E_{x}^{\mathrm{ST}}(\mathbf{x})&=i\!\int\!\dd{\omega}\frac{k_{z}}{k_{\mathrm{sp}}^{2}}\pdv{x}E_{z}^{\mathrm{cos}}(\mathbf{x}\:|\: \omega)\label{Eq:STSPP_x}\\
E_{y}^{\mathrm{ST}}(\mathbf{x})&=i\!\int\!\dd{\omega}\frac{k_{z}}{k_{\mathrm{sp}}^{2}}\pdv{y}E_{z}^{\mathrm{cos}}(\mathbf{x}\:|\: \omega)\label{Eq:STSPP_y}\\
E_{z}^{\mathrm{ST}}(\mathbf{x})&=\!\int\!\dd{\omega} E_{z}^{\mathrm{cos}}(\mathbf{x}\:|\: \omega)\label{Eq:STSPP_z}.
\end{align}
where $k_{z}\!=\!\sqrt{k_{\mathrm{l}}^{2}\epsilon_{\mathrm{m}}-k_{\mathrm{\mathrm{sp}}}^{2}}$ is the wavenumber in the surface normal that determines the exponential decay in the $z$-direction $(\mathbf{E}(x,y,z;t) = \mathbf{E}(\mathbf{x})e^{ik_{z}z}$). Remind that $k _ z$ and $k_{\mathrm{\mathrm{sp}}}$ are functions of $\omega$.

While the $k_{x}$ of the conventional SPP wave packet is always 0, the $k_{x}$ of the ST-SPP wave packet varies with frequency to keep $\dv*{\omega}{k_{y}} = c \tan \theta$. As a result, ST-SPP wave packets possess a finite transverse bandwidth $\Delta k_{x}$ and thus exhibit spatial confinement in both the axial and transverse directions. Previous experiments on ST wave packets in free space and simulation results of the ST-SPP wave packets have shown that the spatial distributions of the ST wave packets are X-shaped. [see Fig.~\ref{Fig:Efield}(f)]~\cite{Yessenov22AOP, Saari97PRL, Ichiji22arXiv}. 
The ST-SPP wave packet with a characteristic X-shaped spatial distribution propagates toward the $y$-direction with the designed group velocity $\widetilde{v_{g}}\!=\!c\,\mathrm{tan}\,\theta$, maintaining the identical spatial distribution unless ohmic losses are considered~\cite{Schepler20ACSP}.

\section{Energy flow density}
\begin{figure*}[t]
\centering
\includegraphics[width=0.9\textwidth]{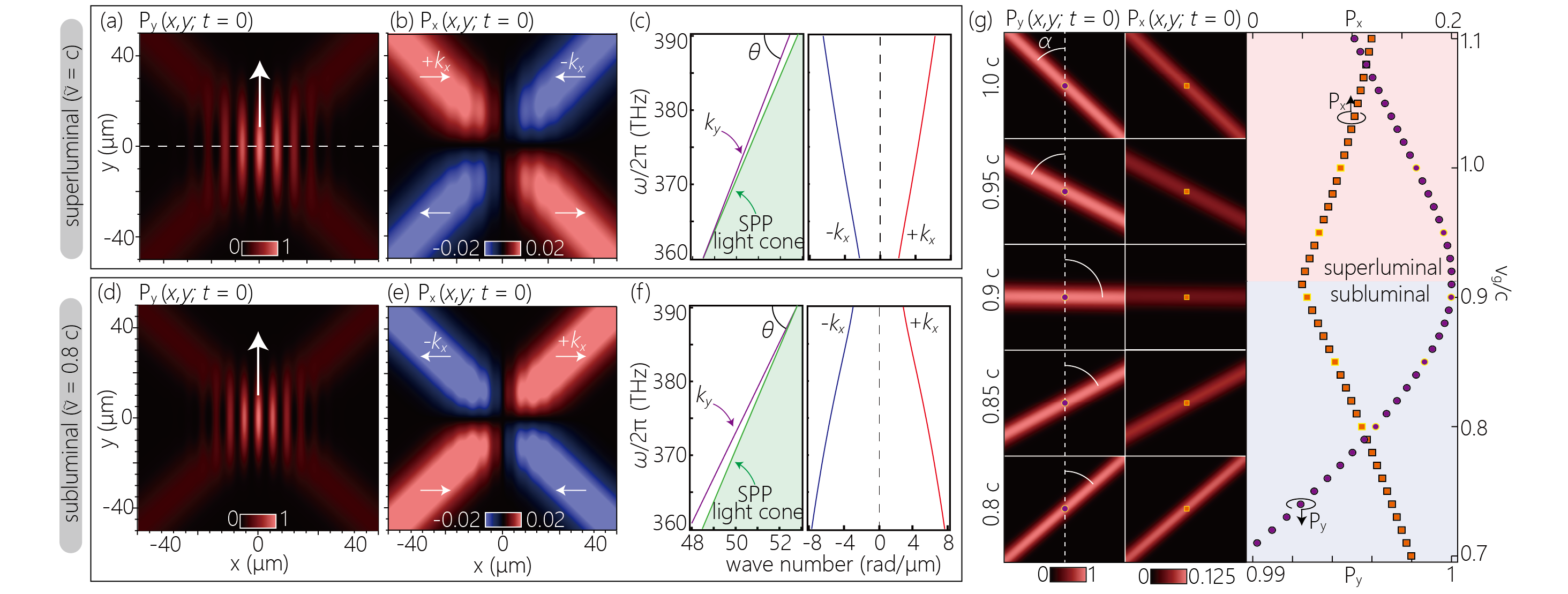}
\caption{The spatial distributions of the energy flow density for ST-SPP waves at $t = 0$. (a) $P_{y}$ and (b) $P_{x}$ components of the superluminal ST-SPP wave packet with $\widetilde{v_{g}}=\!=\!1.0 c$. (c) The spectral projections onto the ($k_{y}$,$\omega$) and ($k_{x}$,$\omega$)-plane. (d-f) Same as (a-c) but for the subluminal ST-SPP wave packets with $\widetilde{v_{g}}=\!=\!0.8 c$. (g) spatial distributions of the energy flow density of one of the branches comprising the ST-SPP wave packet calculated from the $+k_{x}$ components only. All of each display region is identical to (b,e). The intensities obtained at the $x\!=\!y\!=\!0$ are plotted in the right panel as a function of group velocity. The maximum value of $P_{y}$ at $\widetilde{v_{g}}\!=\!v_{\mathrm{sp}}$ was normalized to 1.}
\label{Fig:Poynting}
\end{figure*}

For an SPP wave with an evanescent field, the SAM can be calculated as follows,
\begin{equation}\label{Eq:SAM}
\bm{S}=\frac{1}{4\omega}\mathrm{Im}(\epsilon\bm{E^*}\times\bm{E}+\mu\bm{H^*}\times\bm{H}).
\end{equation}
The electric field $\bm{E}$ is given by Eq.~(\ref{Eq:STSPP_x}-\ref{Eq:STSPP_z}), and the magnetic field $\bm{H}$ is calculated from $E_{z}$ by employing the Maxwell equations~\cite{Shi21PNAS}(See appendix). The asterisks signify their complex conjugates.
We can confirm the SAM is a divergenceless vector field, $\nabla \cdot \bm{S} = 0$.
This implies that the SAM is the rotation of another vector field.
Indeed, taking the rotation of the canonical decomposition of momentum density,
$\nabla \times \epsilon \mu (\bm{P} - \bm{P} _ o) = \nabla \times \nabla \times \bm{S}/2$,
where $\bm{P} = \operatorname{Re} (\bm{E} ^ * \times \bm{H})/2$ is the total energy flow density,
and $\bm{P} _ o = \operatorname{Im}[\mu ^ {-1} \bm{E} ^ * \cdot (\nabla) \bm{E} + \epsilon ^ {-1} \bm{H} ^ * \cdot (\nabla)\bm{H}]/(4\omega)$ is the orbital contribution,
we can find the following relationship~\cite{Bekshaev07OC,Berry09JOA,Bekshaev11JO,Bekshaev13JO,Shi21PNAS}:
\begin{equation}\label{Eq:rot}
\bm{S} = \frac{2}{\omega^2}\nabla\times(\bm{P}-\bm{P} _ o) = \frac{1}{2\omega^2}\nabla\times\bm{P}.
\end{equation}
From Eq.~(\ref{Eq:rot}), it can be confirmed that the SAM of the SPP is represented by a rotation of the energy flow density. 
In Fig.~\ref{Fig:Poynting}, we plot the in-plane components of the energy flow density in the ST-SPP wave packets calculated at different group velocities. In this study, the group velocity of the conventional SPP wave packet at the center frequency, $v_{\mathrm{sp}} (\omega/(2\pi)\!=\!375~\mathrm{THz})\!=\!0.91 c$, is defined as the boundary velocity for superluminal ($\widetilde{v_{g}}\!>\!v_{\mathrm{sp}}$) and subluminal ($\widetilde{v_{g}}\!<\!v_{\mathrm{sp}}$) ($c$: speed of light in vacuum).
The tilt angle of the spectral plane $\theta$ is determined by the designed group velocity, $\theta\!=\!\mathrm{tan}^{-1}(\widetilde{v_{g}}/c)$. In this paper, we set $\omega_{o}/(2\pi) = 360\ \mathrm{THz}\ (390\ \mathrm{THz})$ for the superluminal (subliminal) ST-SPP wave packet. The spectral projections onto the ($k_{y}$,$\omega$)- and ($k_{x}$,$\omega$)-planes for an ST-SPP with $\widetilde{v_g} = c$, i.e., the superluminal ST-SPP wave packet, are plotted in Fig.~\ref{Fig:Poynting}(c).

The energy flow density in the axial direction $P_{y}$ indicates the direction in which the ST-SPP wave packet propagates. The spatial distribution of $P_{y}$ shown in Fig.~\ref{Fig:Poynting}(a) corresponds to the spatial distribution of the field intensity profile shown in Fig.~\ref{Fig:Efield}(f).
In contrast to $P_{y}$, which is positively valued everywhere, the spatial distribution of $P_{x}$ (i.e., the energy flow density in the direction transverse to the propagation axis) is composed of two diagonal branches with opposite signs [see Fig.~\ref{Fig:Poynting}(b)]. These two branches with opposite signs are constructed by the $+k_{x}$ and $-k_{x}$ components of the ST-SPP, indicated by red and blue lines in Fig.~\ref{Fig:Poynting}(c), respectively. As a result of the destructive interference between the two branches, $P_{x}$ vanishes on the $x$- and $y$-axes (at $t\!=\!0$.).
Figs.~\ref{Fig:Poynting}(d-f) show the Poynting vector and the spectral projections for the subluminal ST-SPP wave packet, $\widetilde{v_{g}} = 0.8 c$. In the subluminal and superluminal ST-SPP wave packets, the energy flow in the propagation direction $P_{y}$ exhibits a similar spatial distribution. In contrast, the energy flow in the transverse direction $P_{x}$ is reversed (Fig.~\ref{Fig:Poynting}(e)).

In Fig.~\ref{Fig:Poynting}(g), the calculation for one branch consisting only of the +$k_{x}$ components [red lines in Fig.~\ref{Fig:Poynting}(c,f)] is presented, which is useful to understand the spatial distribution of the ST-SPP wave packet and the energy flow. 
The typical spatial distributions of $P_{y}$ and $P_{x}$ of $+k_{x}$ branches are plotted in the left panels. 
While $P_{x}$ and $P_{y}$ are positive regardless of the group velocity, the spatial distribution of the branches showed a marked dependence on the group velocity. In the superluminal regime, the angle between the branches and the propagation axis is positive, $\alpha\!>\!0$, while, in the subluminal regime, $\alpha\!<\!0$. This implies the counterintuitive fact that the spatial angle of the branch is determined independently of the direction of the energy flow (See appendix).
In addition, the intensity distributions also show that $P_{y}$ and $P_{x}$ have different dependencies on the group velocity. $P_{y}$ and $P_{x}$ at $(x,y)\!=\!(0,0)$ for various group velocities ranging from 0.7 $c$ to 1.1 $c$ are plotted in the right panel. $P_{y}$ reaches the maximum value when $\widetilde{v_{g}}\!=\!v_{\mathrm{sp}}$ and decreases as the $\widetilde{v_{g}}$ deviates from $v_{\mathrm{sp}}$, and vice versa for $P_{x}$. 
Since the ratio of $k_{x}$ and $k_{y}$ represents the tilt angle of each frequency component, an increase in $k_{x}$ tends to result in an increase in $P_{x}$ in exchange for a decrease in $P_{y}$.
It is also noteworthy that the variation of $P_{y}$ is less than 1 $\%$ while that of $P_{x}$ exceeds $50 \%$ in this calculation range. This correlation between $P_{y}$ and $P_{x}$ suggests that the controllable group velocity of the ST-SPP wave packet can be interpreted as the projection of the velocity of the diagonal branch onto the propagation axis, which is propagating in a slightly oblique direction from the propagation axis. 
Note that the velocity projected on the propagation axis corresponds to the velocity of movement of the intersection point of the two branches for the ST-SPP wave packet consisting of $\pm k_{x}$ components.

\section{Spin angular momentum}
\begin{figure*}[t]
\centering
\includegraphics[width=0.9\textwidth]{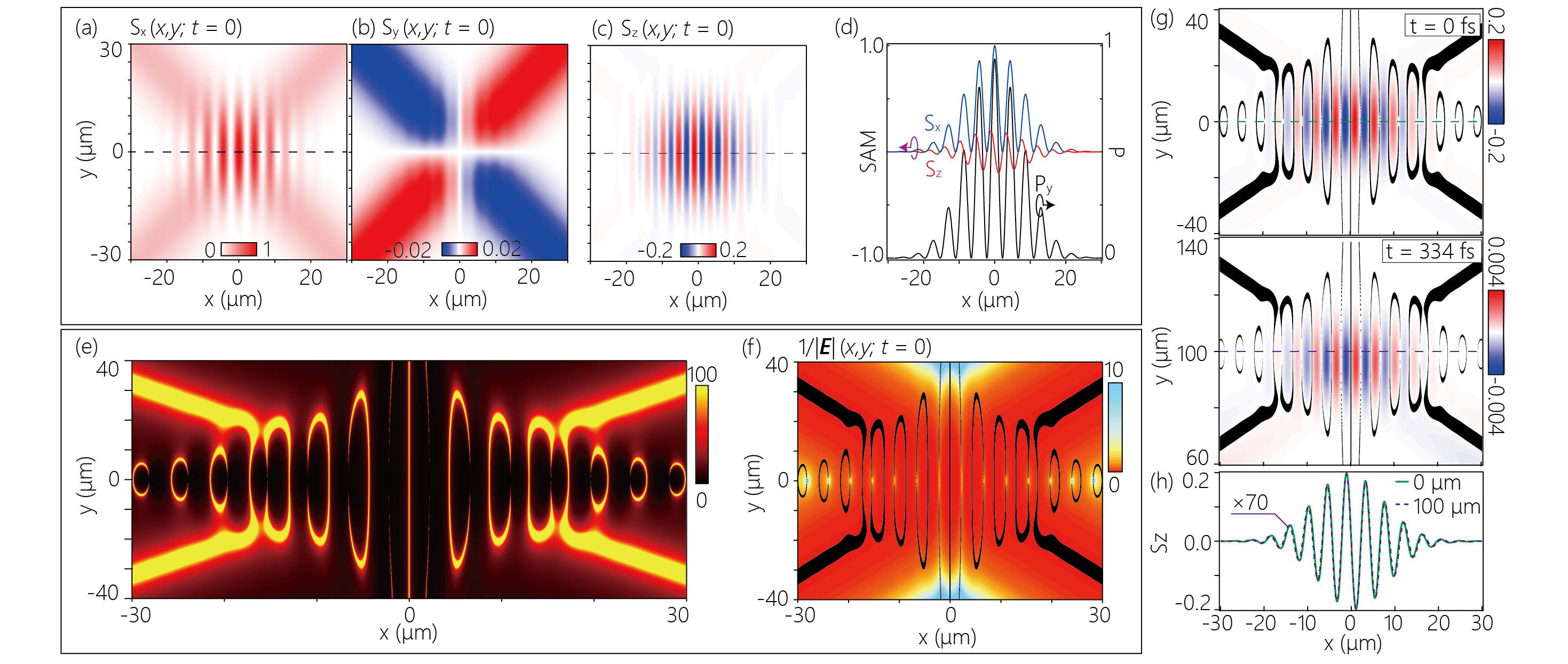}
\caption{(a-c) Spatial field distribution of the spin texture, (a) $S_{x}$, (b) $S_{y}$, and (c) $S_{z}$ component. (d) Cross-sections along the dashed lines in (a,c) and energy flow density in the propagation direction, $P_{y}$, obtained at corresponding positions. (e) The L-line singularity map of the in-plane polarization of SPP fields. (f) The inverse of the magnitude of the electric field distribution. The black lines represent the L-line calculated by taking the contour lines of (d) with a value of 100. 
(g) spatial distributions of the $t$-SAM$_{\perp}$ and L-lines before and after propagation calculated at $t\!=\!0$ and $t\!=\!334$ fs. The cross sections along the lateral direction at $x\!=\!0~\mathrm{\mu}$m and $100~\mathrm{\mu}$m are plotted in (h).
}
\label{Fig:SAM}
\end{figure*}
Figures~\ref{Fig:SAM}(a-c) show each component of the SAM of the superluminal ST-SPP wave packet, calculated from Eq.~(\ref{Eq:SAM}) at $\widetilde{v_{g}} = c$. As the main component of $t$-SAM$_{\parallel}$ ($S _ x$ and $S _ y$) arises from the exponential decay of the in-plane energy flow density along the surface normal, we have $S_{x} \approx -[1/(2\omega ^ 2)]\partial P_{y}/\partial z$ and $S _ y \approx [1/(2\omega ^ 2)]\partial P_{x}/\partial z$ and can confirm these are consistent with the spatial distribution of $P_{y}$ and $P_{x}$ [see Figs.~\ref{Fig:SAM}(a,b)].

On the other hand, the $t$-SAM$_{\perp}$ ($S_{z}$) arises from the rotation of the in-plane energy flow of the SPP. As the $y$-component of the energy flow is approximately 10 times larger than the $x$-component, $P_{y} \gg P_{x}$, in this study [Fig.~\ref{Fig:Poynting}(g)], it dominantly contributes to the $t$-SAM$_{\perp}$, $S_{z}\approx [1/(2\omega ^ 2)]\partial P_{y}/\partial {x}$.
Figure~\ref{Fig:SAM}(d) shows the cross-sections of the $x$-and $z$-components of the SAM along the $x$-axis ($y\!=\!0$) [dashed lines in Figs.~\ref{Fig:SAM}(a,c)]. Note that the $y$-component of the energy flow (dashed line in Figs.~\ref{Fig:Poynting}(a)) is also plotted in the same figure. The spatial distribution of the phase-matched $S_{x}$ and $\pi/2$-shifted $S_{z}$ with respect to the $P_{y}$ component is consistent with the SAM of the Cosine SPP wave reported previously~\cite{Shi21PNAS}.

To characterize the three-dimensional spin texture of the SAM, we evaluate where the ellipticity of the local polarization of the SAM is minimal in the $xy$ plane.
This region corresponds to the equator of the Poincar{\'e} sphere and can be interpreted as the L-line region where the field is linearly polarized~\cite{Foesel17NJP, Dennis02OCmn}. In Fig.~\ref{Fig:SAM}(e), we plotted the L-line singularity map $|\bm{S}|/S_{z}$. 
Note that the L-lines shown in Fig.~\ref{Fig:SAM}(e) are not a one-dimensional trajectory but have a finite width because the L-line map saturates in the region where $S_z$ is smaller than $1\%$.

Because the $t$-SAM$_{\perp}$ of the ST-SPP wave packet is generated from the cosine-like spatial structure resulting from the interference between the $+k_{x}$ and $-k_{x}$ branches, there are almost no $S_{z}$ components in regions where the two branches do not spatially overlap. Therefore, the X-shaped region without fringes of the ST-SPP wave packet exhibits linear polarization, which is the polarization state of the conventional SPP wave packet.
In contrast, multiple split ring-shaped L-lines, which are derived from the spatial structure of the ST-SPP wave packet, are aligned along the $x$-axis ($y\!=\!0$) with their split points oriented in the opposite direction around the $y$-axis.
The split ring structures define perimeters where the $t$-SAM$_{\perp}$ of the ST-SPP wave packet is distributed on the plane, and the split points on the $x$ axis are singularities where the electric field amplitude is zero. Around these discontinuous points, there is a steep gradient in the electric field intensity. In Fig.~\ref{Fig:SAM}(f), we plotted the L-line obtained from Fig.~\ref{Fig:SAM}(e) overlaid on the inverse of the electric field distribution, $1/\bm{|E|}$. 

Although the calculations up to this point only considered the case $t\!=\!0$ and propagation loss was not taken into account, the actual propagation of an ST-SPP wave packet on the dispersive metal surface has a frequency-dependent attenuation determined by the imaginary part of the dielectric function. Attenuating the overall intensity is inevitable, and there will be a slight deformation  where the ST-SPP wave packet has an ultrabroadband spectral width~\cite{Schepler20ACSP,Ichiji22ACSP}. 
To estimate the spin texture stability of the ST-SPP wave packet, calculations of the ST-SPP wave packet before and after propagation were performed with an attenuation model that introduced propagation loss into Eq.~(\ref{Eq:STSPP_x}-\ref{Eq:STSPP_z}). 
The propagation loss at each frequency was calculated from $\mathrm{Im}[\epsilon_{\mathrm{m}}(\omega)]$. 
For each frequency component with a finite transverse wavenumber, which propagates in a slightly oblique direction, we define the distance of propagation $y/\mathrm{cos}(\phi)$, where $\phi$ is the angle between the propagation direction and the $y$-axis, $\phi\!=\!\mathrm{tan}^{-1}(k_{x}/k_{y})$.
In Fig.~\ref{Fig:SAM}(g), we plot the spatial distributions of the $t$-SAM$_{\perp}$ calculated by attenuation model at $t\!=\!0$ and $t\!=\!334 \,\mathrm{fs})$.
The L-lines calculated at each time were also overlaid on the plots.

While the $S_{z}$ distribution after propagation over 100~$\mathrm{\mu}$m shows an overall attenuation in intensity, the spatial distribution surrounded by L-line is almost unchanged from that before propagation. Moreover, the cross-sections along the dashed lines in Fig.~\ref{Fig:SAM}(g) almost overlap [see Fig.~\ref{Fig:SAM}(h)]. This indicates the deformation of the spin texture with propagation is marginal.
From these calculations, we conclude that the ST-SPP wave packet can transport the spin texture in a spatially stable manner on the order of the propagation length of the SPP.

\section{Toplogical charge density}
\begin{figure*}[h!]
\centering
\includegraphics[width=0.9\textwidth]{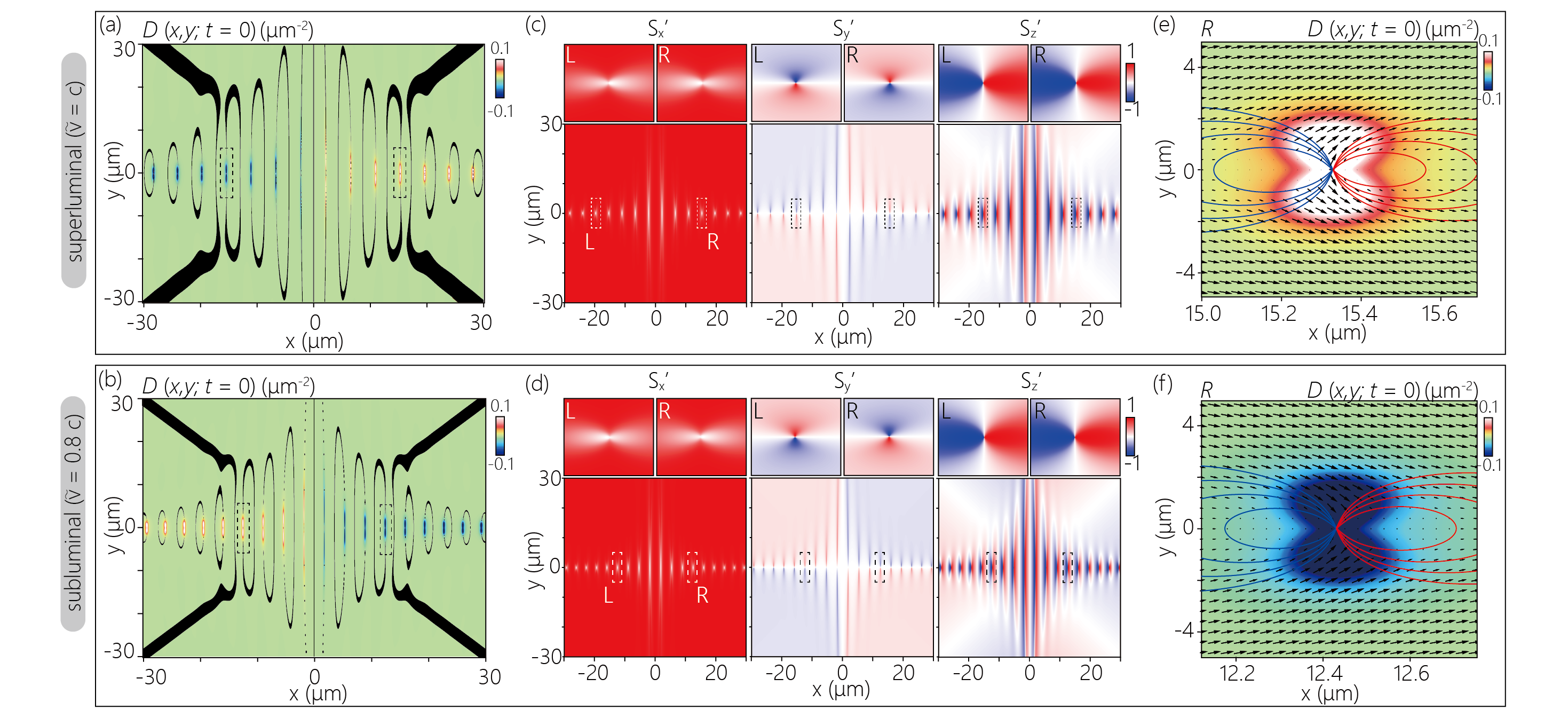}
\caption{(a,b) Spatial distributions of the topological charge density $D$ at $t = 0$ for (a) $\tilde{v} = c$, (b) $\tilde{v} = 0.8  c$. The black lines represent the L-line calculated by $|\bm{S}|/S_{z}$. The regions around the fourth split point on the left and right from the center are defined as $L$ and $R$, respectively. (c,d) Spatial distributions of each normalized SAM component, $S'_{x}$, $S'_{y}$, and $S'_{z}$ for (c) $\tilde{v} = c$ and (d) $\tilde{v} = 0.8 c$. The expanded views of the dashed squares are plotted in the top panels. (e,f) The expanded view of the right dashed square in Figs.~4 (c,d) of the spatial distribution of each component of the normalized SAM. The vector plots show the in-plane components, and the color lines show the contour of the out-of-plane components (red: -0.9 to -0.6, blue: 0.6 to 0.9). The color map represents the distribution of $D$.}
\label{Fig:Dens}
\end{figure*}
We next discuss the topological properties of ST-SPP based on the SAM textures obtained above. We calculate the topological charge density as in studies of topological spin quasiparticles, where the integration of the topological charge density over an area within the boundary of the quasiparticle provides the topological charge (skyrmion number)~\cite{Yanan20Nature, Ghosh23ACSP, Shi21LPR, Zhang21PRR, Lei21PRL, Du19NatPhys, Yanan22APR, Ghosh21APR, Xiong21NL}. The topological charge density is defined by
\begin{equation}\label{Eq:topological}
D=\frac{1}{4\pi}\bm{S'}\cdot\left(\frac{\partial \bm{S'}}{\partial x}\times\frac{\partial \bm{S'}}{\partial y} \right),
\end{equation}
where we have normalised the SAM, $\bm{S'} = \bm{S}/|\bm{S}|$.

Figure~\ref{Fig:Dens}(a) shows the calculated $D$ map for an ST-SPP wave packet with $\widetilde{v_{g}} = c$. The positions of the discretized spatial distribution of $D$ correspond to the split points of the L-lines. One of the major characteristics of the $D$ map is its reversed signs on the left and right sides: the spatial distribution exhibits perfect line symmetry around the $x$-axis, with positive values for positive $x$ and negative values for negative $x$. In addition, the $D$ map for $\tilde{v} = 0.8 c$ plotted in Fig.~\ref{Fig:Dens}(b) shows a similar symmetric spatial distribution as in the case of $\tilde{v} = c$, but with opposite signs on the left and right. The left-right reversal and the group velocity-dependent reversal of the sign of $D$ are explained by the spatial distribution of $S_{y}$. In Figs.~\ref{Fig:Dens}(c,d), we plotted each component of $\bm{S'}$ ($S'_{x}$,$S'_{y}$, and $S'_{z}$) for the superluminal and subluminal ST-SPP waves shown in Figs.~\ref{Fig:Dens}(a,b), respectively. As shown in Fig.~\ref{Fig:SAM}, the $t$-SAM$_{\parallel}$ arising from the decay of evanescent fields has spatial distributions consistent with the in-plane component of $\bm{P}$, $P_{x}$ and $P_{y}$.
While $S'_{x}$ is positive throughout the calculation region regardless of the group velocity, $S'_{y}$ has a different sign in each quadrant. The sign is switched depending on whether the group velocity is subluminal or superluminal.
On the other hand, the spatial distribution of $t$-SAM$_{\perp}$ is mainly determined by the partial differentiation of $P_{y}$ in the $x$-direction; it, therefore, does not exhibit quadrant-dependence of the sign unlike $S'_{y}$. 

To examine the differences in the spatial distribution of each component in detail, the enlarged view around the split points is plotted in the top panels, one on the left (region $L$) and one on the right (region $R$). In contrast to the consistent spatial distribution of $S'_{x}$ and $S'_{z}$, only the $S'_{y}$ component shows that the switched signs for right and left, and for superluminal and subluminal. From Eq.~\ref{Eq:topological}, we can confirm that the sign of $D$ is switched when the two vector components of $\bm{S}'$ are fixed, and only the sign on one component is switched. This is consistent with the calculation results shown in Figs.~\ref{Fig:SAM}(a-d).
It can also be seen that the in-plane and out-of-plane components of $\bm{S}'$ have complementary spatial distributions at the vicinity of the split points. Both $S'_{x}$ and $S'_{y}$ are zero at $y\!=\!0$ and have a spatial distribution that extends toward the $y$-direction from the split point of the L-line. In contrast, $S'_{z}$ shows the maximum value at $y\!=\!0$ and has a distribution that extends toward the $x$-direction. 
The vector plots of the in-plane components of $\bm{S'}$ and the contour plots of the out-of-plane component superimposed on the $D$ map plotted in Figs.~\ref{Fig:Dens}(e,f) show more clearly the positional relationship of each $\bm{S}'$ component. The plotted areas correspond to the region $R$.  
The field distribution of the $S_{x}$ in Figs.~\ref{Fig:Dens}(c,d) and the direction of the in-plane components of the SAM in Figs.~\ref{Fig:Dens}(e,f) indicate that $S_{x}$ is dominant in most of the calculation region. Only in a limited region around the split points of the L-line, each component of SAM has comparable magnitudes, and $D$ has a finite value.

\section{Discussion and Conclusion}
The propagation invariance of an ST-SPP preserves the wave packet shape and the L-line structure, which forms the stable frame of the SAM texture. As discussed above, the propagation of the ST-SPP is accompanied by the propagation of topological charge densities with different signs on its left and right sides. 
In the case of a plasmonic spin skyrmion or meron, a phase singularity (C-point) exists in the core of the vortex where an in-plane rotating electric field (local circular polarization) is formed~\cite{Du19NatPhys, Zhang21PRR, Yanan20Nature, Yanan22APR, Ghosh23ACSP}. It corresponds to the north or the south pole of the Poincar{\'e} sphere; the local polarization is circularly polarized, and the SAM vector points in the surface-normal direction. The field exhibits a $2\pi$-phase rotation around the center with the angular momentum provided by the excitation light field or the geometrical charge of a coupling structure.
The ST-SPP, in contrast, does not have the plasmonic vortex like the plasmonic spin skyrmion; the ST-SPP has an X-shaped electric field distribution, with each branch consisting of the $+k_{x}$ and $-k_{x}$ branch components of the spectral support domain (Fig.~\ref{Fig:Efield}(f), ~\ref{Fig:Poynting}(a-f)). In the central part of the X-shape, the electric field forms a textured structure and oscillates at the central frequency of the wave packet (See appendix).
In the areas on either side, "holes" in the electric field strength are lined up due to the interference between the $+k_{x}$ and $-k_{x}$ branches (Fig.~\ref{Fig:SAM}(f)).
These holes are not accompanied by a $2\pi$-phase rotation of the field and are therefore distinct from the central core of the plasmon skyrmions mentioned above. Because the ST-SPP propagates at the group velocity $v_{\mathrm{sp}}$, the energy flow density is inherently dominated by the positive-valued $P_{y}$ component. The lack of negative $P_{y}$ precludes generations of vortices around phase singularities. However, the periodic modulation of $P_{y}$ intensity in the transverse direction is responsible for the surface-normal SAM (Fig.~\ref{Fig:SAM}(d)). The overall SAM vector distribution of ST-SPP is concentrated in the positive $S_{x}$, while it changes steeply only in the vicinity of  the holes of the field. The SAM vector is not distributed over the hemisphere or the entire Poincar{\'e} sphere, therefore, the topological charge is neither half-integer nor integer. However, the different branches consisting of the $+k_{x}$ and $-k_{x}$ components surrounding the hole generate $S_{y}$ components of different signs, resulting in the winding of SAM and the related nonzero topological charge density $D$ (Fig.~\ref{Fig:Dens}(e,f)). This nonzero $D$ is specific to ultrashort ST-SPP wave packets, not to single frequency SPP fields, because it originates from the spatial proximity of the $+k_{x}$ and $-k_{x}$ branches, and thus should be associated with the geometry of the spectral support domain that generates the ST-SPP.

Since the SPP also has an evanescent field associated with electric polarization on the metal side, it would be useful to describe the SAM interaction with metal electrons and the relation to topological magnetic quasiparticles. Firstly, the SAM in the metal comprises the circular motion of electrons or localized circular currents. As a result, magnetization due to the inverse Faraday effect is expected to occur within the metal~\cite{Bliokh17PRL,deschamps1970inverse,sheng1996inverse,hertel2006theory,zhang2009simple}. The magnetic field can interact with the electron spin.
Indeed, conduction electron spin currents can be driven by the SPPs at a single surface~\cite{oue2020electron, Wijaya2020arxiv, bekshaev2022spin}, a metallic film~\cite{oue2020effects,oue2020optically}, and in graphene~\cite{ukhtary2021spin,tian2022switching}.
Secondly, the SAM vector in the Poincar{\'e} sphere is regarded as an optical correspondence of the magnetization vector in magnetic skyrmion systems. The topological charge density is a physical quantity that reflects the degree of twist in the local spin (magnetization) vector. In the context of magnetic topological spin textures, $D$ can be related to as an emergent magnetic field $b_{\mathrm{em}}$~\cite{Nagaosa13NatN, Gobel21PhysR, Yang21NRP}. In two-dimension systems, the magnitude of the field vector directing the surface normal is proportional to $D$. Importantly, $b_{\mathrm{em}}$ is not merely a fictitious field but can also interact with the spins of other electrons in solid materials. 
The interaction of $b_{\mathrm{em}}$ with electron spins via the magnetic moments described by the Landau-Lifshitz-Gilbert (LLG) equation has been experimentally confirmed~\cite{Hamamoto16APL, Maccariello18NN, Schulz12NP, Jiang17PR, Gobel21PhysR}. This interaction, known as the topological Hall effect~\cite{Bruno04PRL, Neubauer09PRL, Lee09PRL}, has realized the manipulation of magnetic skyrmions by external currents. 
It is noteworthy that the \textit{"propagating topological charge density"} was found in the context of SPPs, which belong to the class of optical topological systems. Furthermore, the correspondence between the magnetic topological systems and optical spin topological systems demonstrated in various previous studies suggest that the topological charge density calculated in this study can be considered as an optical emergent magnetic field that can interact with the spins of electrons in materials~\cite{Jiang20PRL}.
In addition, the propagation invariance and arbitrary group velocities of the ST-SPP wave packet enable adaptive transport of topological charge density with a stable spatial distribution across subluminal and superluminal regimes. Through the experimental excitation of ST-SPP wave packets using a spatial light modulator (SLM), the intensity, spatial distribution, and polarity of $D$ can be adjusted externally at the driving frequency of the SLM.
This feature enables the exploration of novel opportunities to realize spin information transmission.

In conclusion, we have conducted numerical simulations to examine the spin texture in an ST-SPP wave packet, which is a type of plasmonic bullet that possesses propagation invariance at arbitrary group velocities. Our calculations reveal three-dimensional spin textures in these wave packets, which are found to have stable spatial distributions during propagation over distances on the order of the SPP propagation length. Furthermore, we have discovered that the ST-SPP wave packet also possesses a finite topological charge density, which exhibits a symmetric spatial distribution where only the sign differs between the left and right sides of the propagation axis. As the spin texture and topological charge density are determined by the gradient of the energy flow density, it is possible to design them by manipulating the ST-SPP wave packet. These findings suggest that three-dimensional localized electromagnetic fields can propagate with accompanying spin textures and may offer the possibility of observing optical topological particle propagation phenomena experimentally.

\section*{Acknowledgments}
A.K.~thanks Y.~Dai for the valuable discussions.
\section*{Funding Sources}
This work was supported by the JSPS KAKENHI (JP16823280, JP18967972, JP20J21825); MEXT Q-LEAP ATTO (JPMXS0118068681); and by the U.S. Office of Naval Research (ONR) under contracts N00014-17-1-2458 and N00014-20-1-2789.
D.O.~is supported by the President's PhD Scholarships at Imperial College London, by JSPS Overseas Research Fellowship, by the Institution of Engineering and Technology (IET), and by Funda\c{c}\~ao para a Ci\^encia e a Tecnologia and Instituto de Telecomunica\c{c}\~oes under project UIDB/50008/2020.

\appendix
\renewcommand\thefigure{S\arabic{figure}} 
\setcounter{figure}{0}  

\section{The electric and magnetic field}
The electric field brought about by an SPP is given by utilizing the transverse polarisation basis with the SPP dispersion relation.
As the SPP wave is a transverse magnetic mode, the electric field will be given, in terms of the Hertz potential~\cite{nisbet1955hertzian,hacyan1990spectrum,jackson1999classical}, in the reciprocal space as 
\begin{align}
\bm{E} = \bm{k} \times \bm{k} \times \psi \bm{u}_z = (k_z \bm{k} _\parallel - |\bm{k}_\parallel|^2 \bm{u}_z)\psi,
\end{align}
where $\psi\bm{u}_z$ corresponds to the Hertz vector potential, and we defined the parallel component of the wave vector, $\bm{k}_\parallel = k_x \bm{u}_x + k_y \bm{u}_y$.
The ratio of $E_{x,y}$ to $E_z$ can be written as 
\begin{align}
\frac{E_{x,y}}{E_z} = \frac{k_z k_{x,y}}{-|\bm{k}_\parallel|^2} = \frac{ik_z \partial_{x,y}}{k_{\mathrm{sp}}^2}. 
\end{align}
Note that we have used the dispersion relation $|\bm{k}_\parallel|^2 = k_{\mathrm{sp}}^2$ and replaced $k_{x,y} \rightarrow -i \partial_{x,y}$.
Multiplying $E _ z$ from the right-hand side, we can get
\begin{align}
\bm{E} _ \parallel = \frac{ik_z}{k_{\mathrm{sp}}^2} \nabla _ \parallel E_z.
\end{align}
where we have defined the parallel component of the electric field, $\bm{E} _ \parallel = \bm{E} - E _ z\bm{u} _ z$,
and the derivative in the direction parallel to the surface, $\nabla _ \parallel := \bm{u} _ x\partial _ x + \bm{u} _ y\partial _ y$.

Once we get the electric field, the magnetic field can be calculated from Faraday's law of induction,  $\nabla \times \bm{E} = \omega \mu _ 0\bm{H}$, as follows:
\begin{align}
\bm{H} &= \frac{\nabla\times\bm{E}}{\omega \mu _ 0}
= \frac{(\nabla _ \parallel + k _ z \bm{u} _ z) \times (\bm{E} _ \parallel + E _ z \bm{u} _ z)}{\omega \mu _ 0},\\
&= i\frac{1+k _ z ^ 2/k _ \mathrm{sp} ^ 2}{\omega \mu _ 0} \bm{u} _ z \times \nabla _ \parallel E _ z,\\
&= i\frac{\omega \epsilon \epsilon _ 0}{k _ \mathrm{sp} ^ 2}  \bm{u} _ z \times \nabla _ \parallel E _ z.
\end{align}
Note that we have applied $k _ \mathrm{sp} ^ 2 + k _ z ^ 2 = 
(\omega/c)^2 \epsilon$.
Note also that the permittivity is dependent on whether we are on the dielectric or metal side,
\begin{align}
    \epsilon (z,\omega) = \begin{cases}
    \epsilon _ \mathrm{d} & (z > 0),\\
    \epsilon _ \mathrm{m}(\omega) & (z < 0).
    \end{cases}
\end{align}

\section{Wavefront direction}
\begin{figure*}[h!]
\centering
\includegraphics[width=0.9\textwidth]{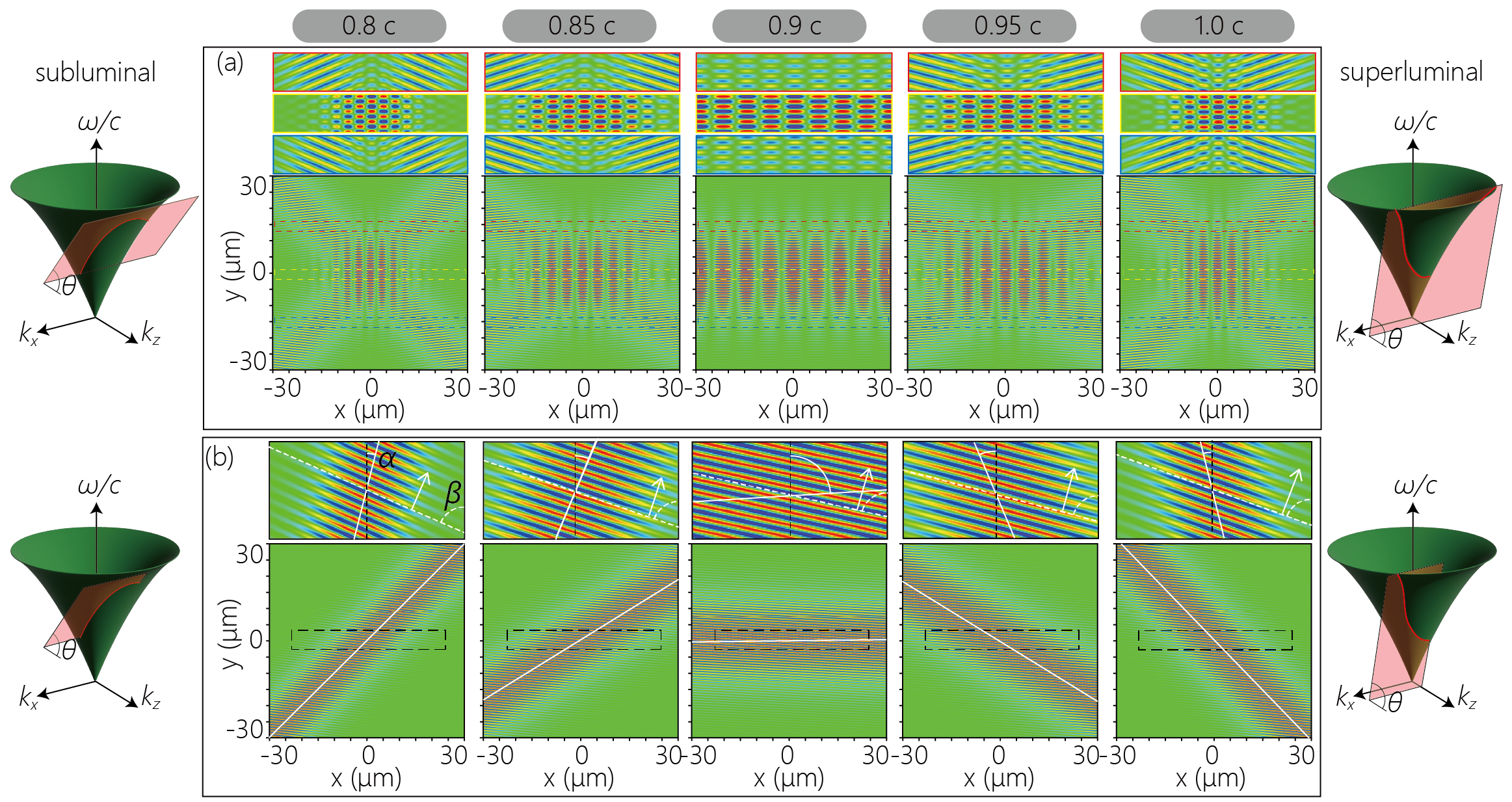}
\caption{(a) Spatial distribution of the ST-SPP wave packet, Re($E_{z} (x,y;t=0)$) for different group velocity ranging from 0.8 $c$ to 1.0 $c$. Top panels show the expanded views indicated by dashed squares. The colors of the frames (red, yellow, and blue) correspond to the colors of the dashed squares in the main panels. (b) One of the two branch structures that make up the ST-SPP wave packet, calculated with only the positive $k_{x}$ component. Top panels show the expanded views indicated by black dashed squares. White solid lines in main panels and dashed lines in top panels indicate the peak positions of band structures itself and wavefronts of inside the band structures. White arrows are normal direction of the wavefront correspond to the propagate directions. The angles between each line and $y$-axis were defined as $\alpha$ and $\beta$. Note that the $\alpha$ and $\beta$ are not comparable quantities since the aspect ratio of the expanded views has been modified.}
\label{Fig:sp1}
\end{figure*}

The electric spatial distribution, including fine fringes within the wave packet, is useful for visualizing the intricate motion of the ST-SPP wave packets. Fig.~\ref{Fig:sp1} displays the real part of the electric field $E_{z}$ $(x,y;t=0)$ of the ST-SPP wave packets calculated for different group velocities ranging from $0.8~c$ to $1.0~c$. The expanded views of the dashed squares are plotted in the top panels.

The entire spatial distributions show that the subluminal and superluminal ST-SPP wave packets with close $v_{\mathrm{diff}}$ have similar envelope distributions, where $v_{\mathrm{diff}}$ is a difference of the group velocity between the ST-SPP wave packet and the conventional SPP wave packet $(v_{\mathrm{diff}}\!=\!|v_{\mathrm{sp}}-\widetilde{v_{g}}|)$. The angles $\pm\alpha$ of the $\pm k_{x}$ branches, which constitute the ST-SPP wave packet, depend on the angle $\theta$ of the spectral plane and vary with the group velocity.
While the entire distribution of the wave packets corresponds, the fringes within the branches are oriented in contrasting directions in the superluminal and subluminal wave packets. The wavefronts of the subluminal ST-SPP are distributed in a rhombic shape surrounding the central region, whereas the wavefronts of the superluminal ST-SPP are distributed radially from the central region.
In Fig.~\ref{Fig:sp1}(b), we plotted the $+k_{x}$ branch calculated from only the $+k_{x}$ component of the ST-SPP wave packet. The angles between the wavefront inside the branch made with the propagation axis are defined as $\beta$. The value of $\alpha$ varies widely with $\widetilde{v_{g}}$, being negative during subluminal and positive during superluminal. In contrast, the group velocity dependence of $\beta$ is relatively small compared to $\alpha$, and the wavefront faces the upper right direction regardless of $v_{g}$. 

\section{Spatial distribution}
\begin{figure*}[h]
\centering
\includegraphics[width=0.9\textwidth]{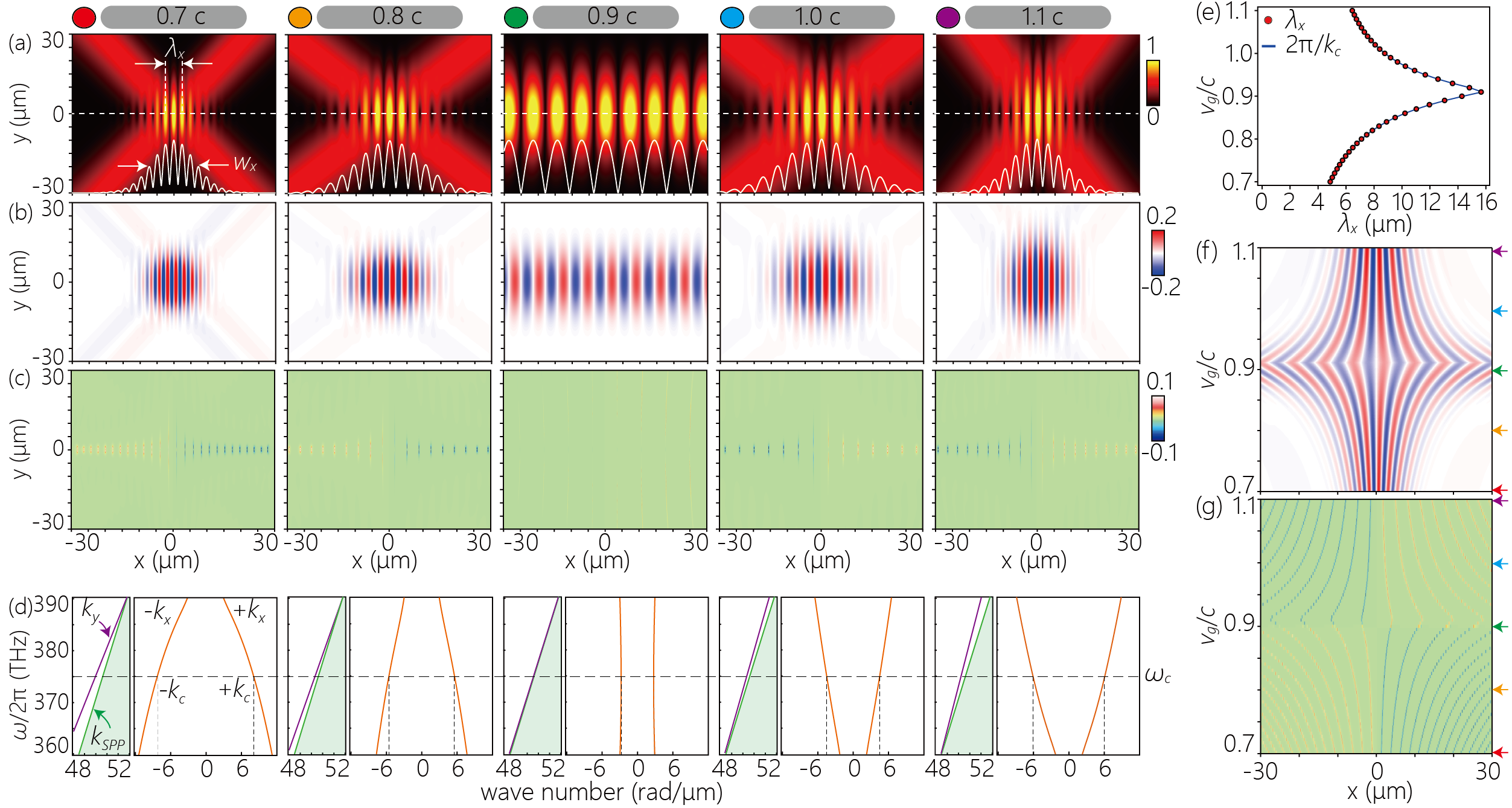}
\caption{(a-c) Spatial field distribution of the typical parameters of ST-SPP wave packets in physical space for various group velocities ranging from $0.7~c$ to $1.1~c$, (a) $I(x,y,; t\!=\!0)$, (b) $t$-SAM$_{\perp}(x,y;t\!=\!0)$, and (c) $D(x,y;t\!=\!0)$. White lines in (a) are cross-sections along $y\!=\!0$. (d) The spectral projections onto the ($k_{y}$,$\omega$) and ($k_{x}$,$\omega$)-plane. (e) Transverse wavenumber of the ST-SPP wave packets. The red circles and blue line represent the $\lambda_{x}$ evaluated from $I(x,y;t\!=\!0)$ (shwon in Fig.~S2 (a)) and $2\pi/k_{c}$ calculated from spectral domain (shown in Fig.~S2 (d)), respectively. (f,g) Variation of the (f) $S_{z}$ and (g) $D$ distribution with group velocity. The horizontal profiles correspond to the integral cross sections along the $y$-direction of the $S_{z}(x,y;t\!=\!0)$ and $D(x,y;t\!=\!0)$ in the calculation region shown in Fig.~\ref{Fig:sp2}(b,c) $(-30\!<\!y\!<\!30) ~\mathrm{\mu}$m.
}
\label{Fig:sp2}
\end{figure*}
As discussed in Fig.~3 and Fig.~4 in the main text, the spatial distribution of the SAM and topological charge density are determined by the spatial distribution of the ST-SPP wave packet, which vary with the designed group velocity. In Fig.~\ref{Fig:sp2}(a-c), we plotted the $I(x,y;t\!=\!0)$, $t$-SAM$_{\perp}(x,y;t\!=\!0)$, and $D(x,y;t\!=\!0)$ calculated for different group velocities ranging from 0.7 $c$ to 1.1 $c$. The projections of the spectral support domains onto the $(k_{y},\omega)$ and $(k_{x},\omega)$ planes at each group velocity were plotted in Fig.~\ref{Fig:sp2}(d).

In both the superluminal and subluminal cases, the spectral domain projected onto the $(k_{y},\omega)$ plane moves away from the SPP dispersion curve as the $v_{\mathrm{diff}}$ increase. Therefore, the transverse wavenumber $k_{x}(\omega)\!=\!\sqrt{k_{\mathrm{sp}}^{2}-k_{y}^{2}}$ increases as $v_{\mathrm{diff}}$ increases, resulting in the transverse fringe density of the ST-SPP wave packet is at its lowest value at $\widetilde{v_{g}} = v_{\mathrm{sp}}$.
Under the calculation conditions of this paper, in which the spectrum of the ST-SPP wave packet is assumed to have a Fourier-limited pulse with a peak at $\omega_{c}\!=\!375$ THz , the transverse wavelength $\lambda_{x}$ of the ST-SPP wave packet is expressed as $\lambda_{x}\!=\!2\pi/k_{c}$ (Fig.~\ref{Fig:sp2}(e)), where $k_{c}$ is transverse wavenumber evaluated at $\omega_{c}$ (Fig~\ref{Fig:sp2}(d)).

Since a region extent of the branches distributed on the $y\!=\!0$ axis is determined by the angle of the branches, the transverse Gaussian spatial distribution of the envelope of the ST-SPP wave packet, $w_{x}$ also vary with $v_{g}$.
This change of the spatial width can also be interpreted as a change in wave packet width due to the difference in $\Delta k_{x}$.
For $\widetilde{v_{g}}\!\simeq\!v_{\mathrm{sp}}$, the spectral domain in $(k_{x},\omega)$ plane shows a slight curvature but almost constant $|k_{x}|$, resulting in very small $\Delta k_{x}$. The spatial distribution of the wave packet is similar to that of a striped ST-SPP wave packet whose spectral domain is defined by iso-$k_{x}$ plane~\cite{Ichiji22ACSP}, and $w_{x}$ is almost infinite.
In contrast, for ST-SPP wave packets with large $v_{\mathrm{diff}}$, the difference between $k_{y}$ and $k_{\mathrm{sp}}$ is larger for larger $\Omega\!=\!\omega_{o}-\omega$, resulting in large $\Delta k_{x}$. From Fig.~\ref{Fig:sp2}, we can confirm the inverse correlation between $\Delta k_{x}$ and $w_{x}$.

A two-dimensional plot of the cross-sections of the $t$-SAM$_{\perp}$ (Fig.~\ref{Fig:sp2}(b)) integrated along the $y$-direction is plotted in Fig.~\ref{Fig:sp2}(f). The vertical axis shows the group velocity normalized by the speed of light in a vacuum.
Since the spatial distribution of the $t$-SAM$_{\perp}$ reflects the intensity profile of the ST-SPP wave packet, the $t$-SAM$_{\perp}$ becomes densely distributed in the central region as $v_{\mathrm{diff}}$ increases. In addition, the maximum intensity of the SAM also increases with increasing $v_{\mathrm{diff}}$ because finer fringes imply a larger gradient of the energy flow density.
A two-dimensional plot of the integrated cross section of $D$ plotted in Fig.~\ref{Fig:sp2}(g) shows a spatial distribution corresponding to the SAM. The left-right reversal of the sign of $D$ associated with the change in $\widetilde{v_{g}}$ discussed in the main text can be confirmed.

\bibliography{sorsamp}

\end{document}